\documentclass[structabstract]{aa}  
\usepackage{graphicx}
\usepackage{sidecap}
\usepackage{txfonts}
\usepackage{natbib}

\begin{document}
   \title{Continuous upflows and sporadic downflows observed in active regions}


   \author{S. Kamio\inst{1}
          \and
           H. Peter\inst{1} \and W. Curdt\inst{1} \and S. K. Solanki\inst{1,2}}

   \institute{Max-Planck-Institut f\"ur Sonnensystemforschung (MPS),
Max-Planck-Str. 2, 37191 Katlenburg-Lindau, Germany\\
\email{skamio@spd.aas.org}
\and
School of Space Research, Kyung Hee University, Yongin, Gyeonggi 446-701, Korea
}

   \date{Received ; accepted }

 
  \abstract
   {}
   {We present a study of the temporal evolution of coronal loops in active regions and its implications for the dynamics in coronal loops.}
   {We analyzed images of the Atmospheric Imaging Assembly (AIA) on the Solar Dynamics Observatory (SDO) at multiple temperatures to detect apparent motions in the coronal loops.}
   {Quasi-periodic brightness fluctuations propagate upwards from the loop footpoint in hot emission at 1~MK, while sporadic downflows are seen in cool emission below 1~MK.
The upward motion in hot emission increases just after the cool downflows.
}
   {The apparent propagating pattern suggests a hot upflow from the loop footpoints, and is considered to supply hot plasma into the coronal loop, but a wavelike phenomenon cannot be ruled out.
Coronal condensation occasionally happens in the coronal loop, and the cool material flows down to the footpoint.
Emission from cool plasma could have a significant contribution to hot AIA channels in the event of coronal condensation.}


   \maketitle
%

\section{Introduction}

Heating of a coronal loop is a fundamental problem
in solar physics.
Coronal loops, which are thin thread-like structures seen in
coronal emission lines, 
must be continuously heated to account for their radiative
and conductive energy losses.
However, models of coronal loops
are not mature enough to reproduce the observations sufficiently
\citet{winebarger2002},\citet{warren2006}, and \citet{aschwanden2009}
show that a static loop model is not consistent with observations.
A dynamic model with a collection of impulsive heatings
is in better agreement with observations,
but still not satisfactory
\citep{warren2003,warren2007}.
A detailed study of the temporal evolution of coronal loops
is therefore crucial for understanding them.

Observations of nonflaring active regions show that
coronal loops are visible over a wide temperature range and
are highly variable.
Brightness distribution of EUV emission lines indicates that
hot loops ($\geq 2\times 10^6$~K) are concentrated in the cores of active
regions, while lower temperature loops ($\sim 1\times 10^6$~K) are located
on the periphery of the active region
\citep{aschwanden2008c,tripathi2008,odwyer2011}.
Loop-like structures are also seen at
transition-region and chromospheric temperatures.
\citet{kjeldsethmoe1998} have shown that cool loops seen at
$1\sim5 \times 10^5$~K
change significantly within one hour,
while hot loops are less variable.
\citet{schrijver2001} finds cool plasma ($\leq1\times 10^5$~K)
flowing down along the loop, which he interprets as
a result of catastrophic cooling at the loop top.
Numerical simulations shows that frequent occurrence of
falling blobs in chromospheric lines, or coronal rain,
favors a heating concentrated near the footpoints of
a coronal loop \citep{mueller2003,mueller2004,mueller2005,antolin2010}.
Clearly, in order to understand the energy balance of coronal loops,
it is important to study both the cooling and
heating processes.

The Atmospheric Imaging Assembly \citep[AIA;][]{lemen2011,boerner2011} on
the Solar Dynamics Observatory (SDO)
records high-cadence EUV images at multiple temperatures.
In this paper, we selected nonflaring active regions at the limb and
on the disk to study the temporal evolution of coronal loops.
Unprecedented data from AIA allowed studying the flows along the loops
over a wide temperature range.

The paper is organized as follows.
Observations with AIA are described in Sect.~2.
Section 3 shows the temporal evolution of selected loops in the active region
and simulated lightcurves of cooling plasma under constant pressure and
Sect~4 summarizes the results and discusses
the implications for coronal loop models.


\begin{figure*}
 \centering
 \includegraphics[width=16cm]{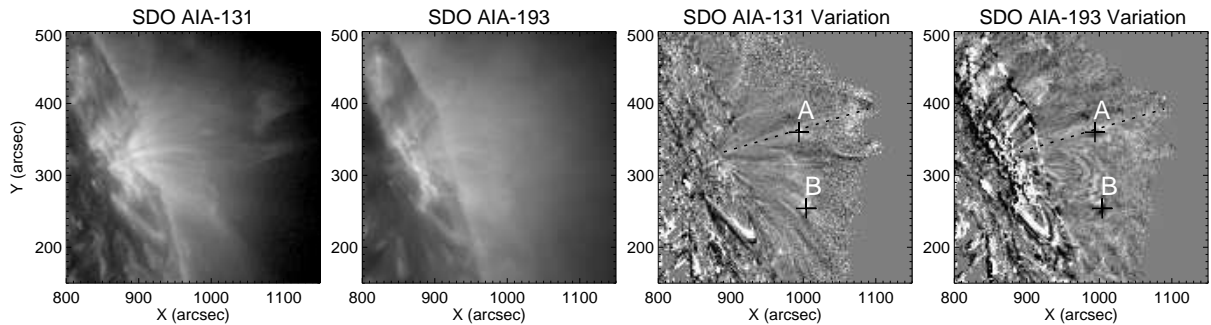} 
 \caption{Active region NOAA 11117 at the limb
as seen in AIA-131\AA\, and 193\AA\, channels on 31 October 2010.
The two right panels show temporal variations between 13:15 UTC and
13:25 UTC (See Sect. 3.1).
The cross signs {\it A} and {\it B} indicate darkening in
AIA-193\AA\, coinciding with brightenings in AIA-131\AA.}
 \label{fig_limb}
\end{figure*}

\begin{SCfigure*}
 \includegraphics[width=14cm]{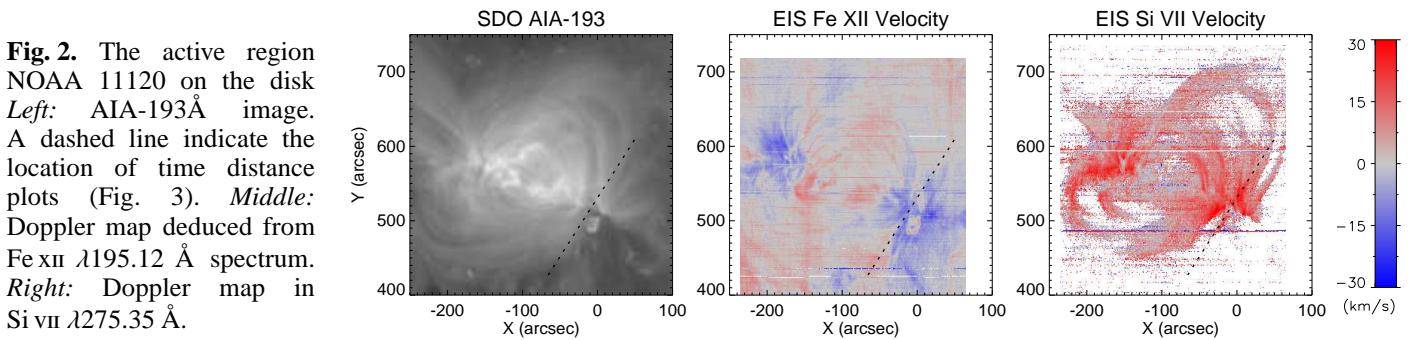} 
 \caption{The active region NOAA 11120 on the disk {\it Left:} AIA-193\AA\,
image. A dashed line indicate the location of time distance plots (Fig. \ref{fig_slice_m}).
{\it Middle:} Doppler map deduced from \ion{Fe}{xii}~$\lambda$195.12~\AA\, spectrum.
{\it Right:} Doppler map in \ion{Si}{vii}~$\lambda$275.35~\AA.}
 \label{fig_diverge}
\end{SCfigure*}

\section{Observations}

AIA records full-Sun images with 0.6\arcsec\, spatial sampling
and 12~s temporal cadence, and
covers all active regions visible from Earth.
We mainly analyzed the following AIA channels to pick up
different temperatures:
304\AA\, (\ion{He}{ii},  $5\times10^4 \mathrm{K}$),
131\AA\, (\ion{Fe}{viii}, $4\times10^5 \mathrm{K}$),
171\AA\, (\ion{Fe}{ix}, $7\times10^5 \mathrm{K}$),
193\AA\, (\ion{Fe}{xii}, $1\times10^6 \mathrm{K}$), and
335\AA\, (\ion{Fe}{xvi}, $3\times10^6 \mathrm{K}$).
The primary ions in the AIA channels are adopted from
\citet{lemen2011}.
Flare lines ($> 10^7 \mathrm{K}$) are excluded
since they are negligible in a nonflaring active region.
A six-hour series of AIA data was extracted from
the calibrated level-1 data archive to study
the active region NOAA 11117 near the limb
on 31 October 2010 (Fig. \ref{fig_limb}).

Another five-hour series of the active region NOAA 11120
crossing the meridian on 5 November 2010 is selected to provide
a different view angle.
At that time, the EUV Imaging Spectrometer \citep[EIS;][]{culhane2007}
on {\it Hinode} \citep{kosugi2007} recorded spectra by employing a 1\arcsec\,
wide slit with 30 s exposure time.
A scan of 300\arcsec$\times$400\arcsec\,  took 158 min.
The calibration of EIS data including dark current subtraction
and removal of hot pixels is performed by using a standard procedure
provided in the Solar Software tree \citep[SSW;][]{freeland1998}.
The Doppler shifts of \ion{Fe}{xii}~$\lambda$195.12~\AA\,
and \ion{Si}{vii}~$\lambda$275.35~\AA\,
are deduced by fitting their spectra with single Gaussians
and by applying the calibration
procedure described in \citet{kamio2010b}.
The rest positions of spectra are determined by
spectra at the limb obtained on 1 November 2010 using
the same observing sequence of EIS.

\section{Results}

\subsection{Continuous propagating pattern}

\begin{figure}
 \centering
 \includegraphics[width=6cm]{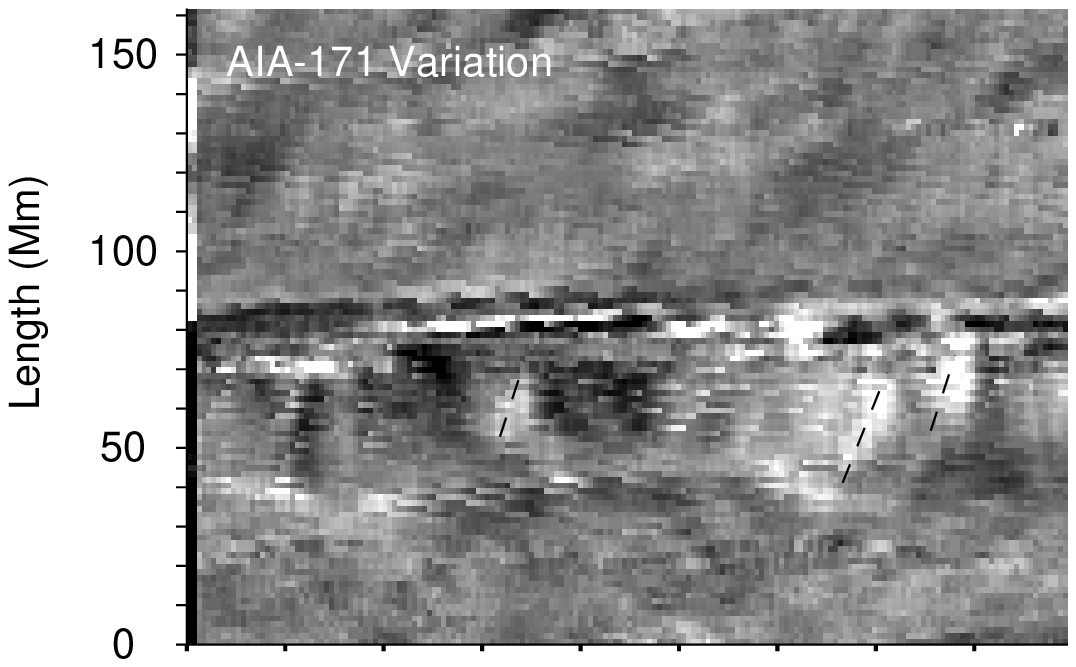}
 \includegraphics[width=6cm]{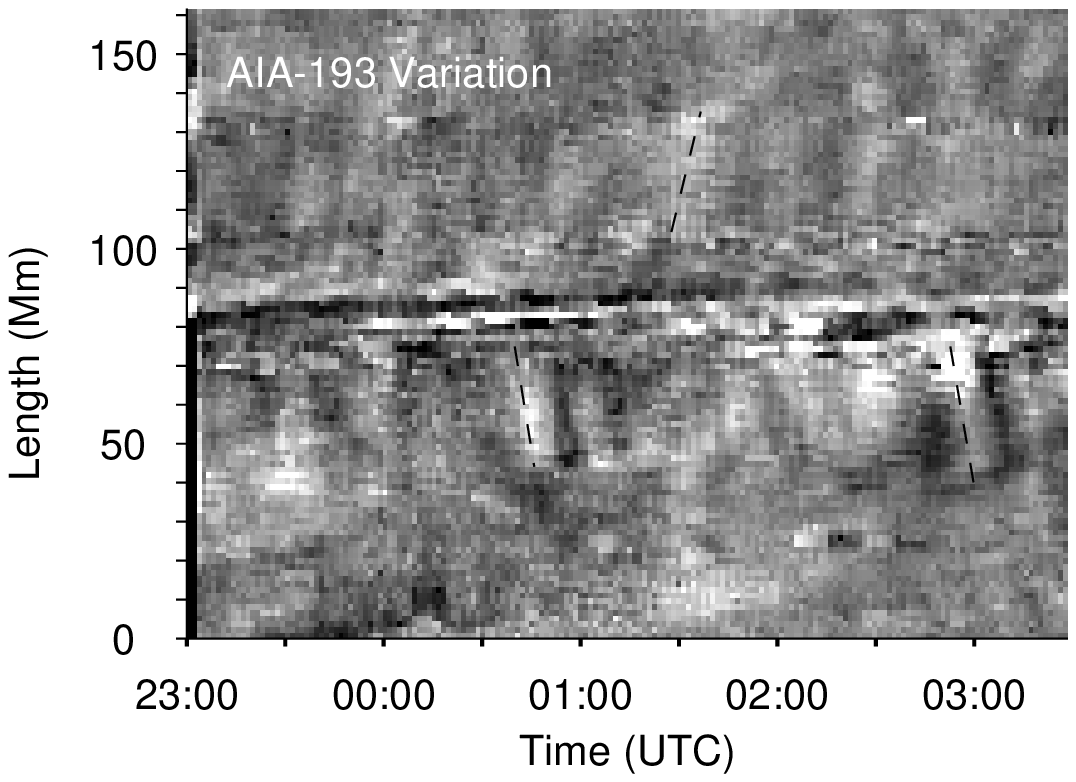}
 \caption{Time distance plots
of AIA-171\AA\, and 193\AA\, temporal variations
along the cut in Fig. \ref{fig_diverge}
displayed on a scale ranging from $-20\%$ to $+20\%$ (black to white).
Dashed lines on AIA-171\AA\, streaks correspond to the
speed of 35 to 43~km~s$^{-1}$.
The apparent velocities of lines on AIA-193\AA\, panel are 57 to 88~km~s$^{-1}$.
}
\label{fig_slice_m}
\end{figure}

\begin{figure}
 \centering
 \includegraphics[width=8cm]{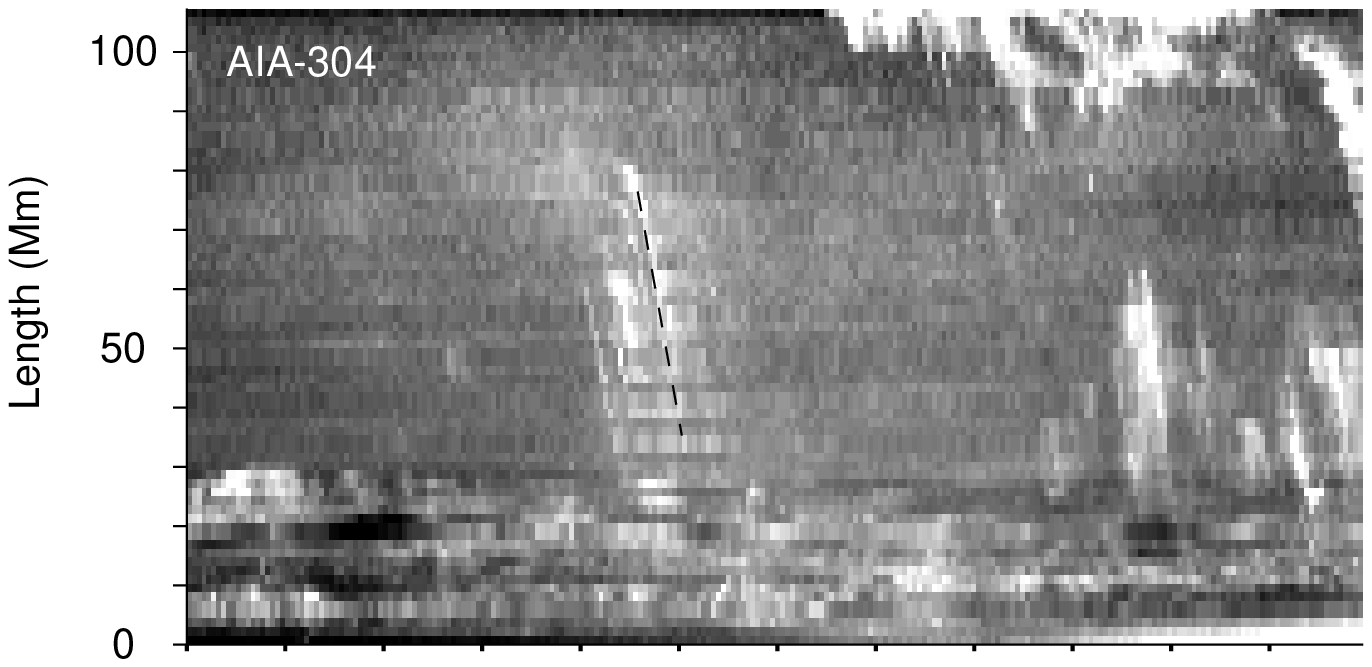}
 \includegraphics[width=8cm]{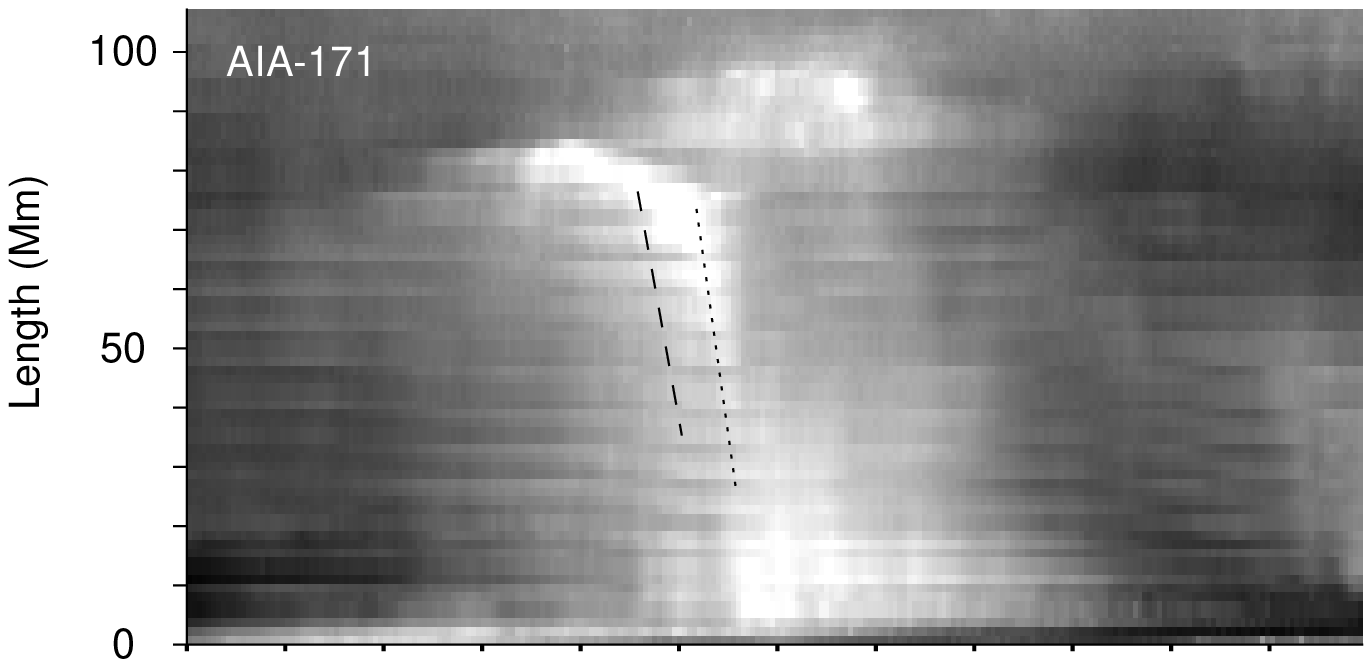}
 \includegraphics[width=8cm]{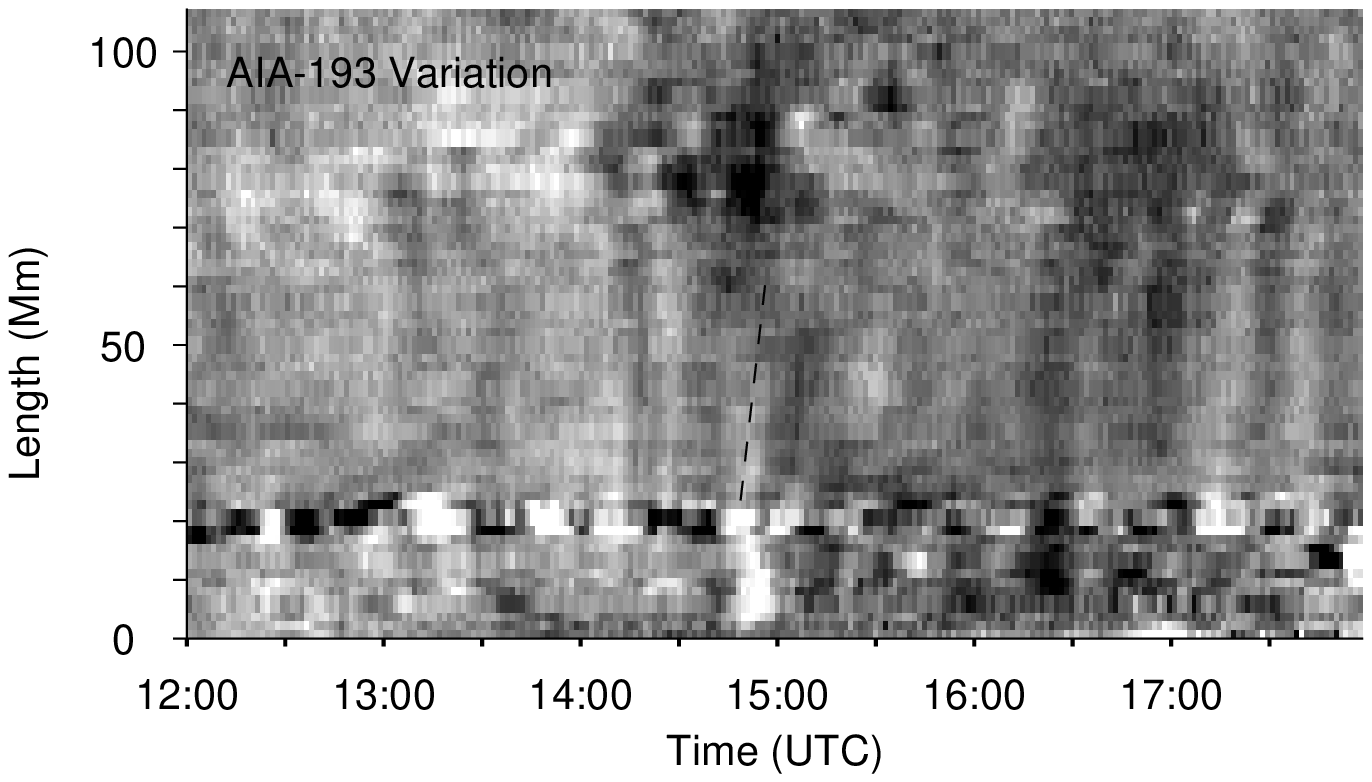}
 \caption{Time distance plots of AIA-304\AA, 171\AA, and temporal variation of 193\AA\, along the dashed line in Fig. \ref{fig_limb}.
The vertical axis shows the distance from the bottom of the slice.
A dashed line on AIA-304\AA\, traces a falling blob, which is also overplotted
in the AIA-171\AA\, panel.
The following falling blob in AIA-171\AA\, is marked by the dotted line.
The bottom panel shows a temporal evolution of AIA-193\AA\,
displayed on a scale from $-8\%$ to $+8\%$.}
 \label{fig_slice}
\end{figure}

An AIA-193\AA\, image and Doppler maps in two spectral lines
from EIS scans of NOAA 11120 are plotted in Fig. \ref{fig_diverge}.
The shifts in \ion{Fe}{xii} indicate
upflows of up to $-20$~km~s$^{-1}$ at the footpoint of the loop system,
while \ion{Si}{vii}, formed at $6\times10^5$~K, exhibits
downflows of up to 30~km~s$^{-1}$.
A temporal variation is computed from a pair of AIA images recorded ten minutes apart,
\begin{equation}
\frac{dI}{I} = \frac{I_2 - I_1}{I_1} = \frac{I_2}{I_1} - 1
\end{equation}
where $I_1$ and $I_2$ are the radiance from the image pair. We employ $dI/I$ instead of $dI$ to highlight relative changes,
which are appropriate since the radiance displays a wide
range of values over the field of view.
Figure~\ref{fig_slice_m} displays time distance plots of
AIA-171\AA\, and 193\AA\, temporal variations
along the line in Fig. \ref{fig_diverge}.
The slice is chosen to cross the footpoint of the loop system.
In the footpoint region between 70~Mm and 90~Mm,
the time slice plots only indicate long-term variation.
Outside the footpoint region, the AIA-193\AA\, exhibits an oppositely
directed quasi-periodic pattern, present throughout the data series,
indicating motions away from the footpoint.

Occasionally, bright streaks moving towards the footpoint
are observed in the AIA-171\AA\, time distance plot, which are
indicated on the plot.
The apparent velocities are 35 to 43~km~s$^{-1}$.
The nonuniform and variable background emission hinders
any detection of such a pattern in AIA-131\AA\, and AIA-304\AA.
When the bright streak in AIA-171\AA\, reaches the footpoint,
a noticeable propagation in AIA-193\AA\, is initiated from the footpoint.
The dashed lines in AIA-193\AA\, plot mark the distinct events that occur
right after the AIA-171\AA\, downflow event.
The magnitude of the outward velocities vary
from 57~km~s$^{-1}$ to 88~km~s$^{-1}$.
The outward streaks show quasi-periodic pattern of about 10 minutes.

\subsection{Sporadic downflow}

\begin{figure}
 \centering
 \includegraphics[width=0.95\columnwidth]{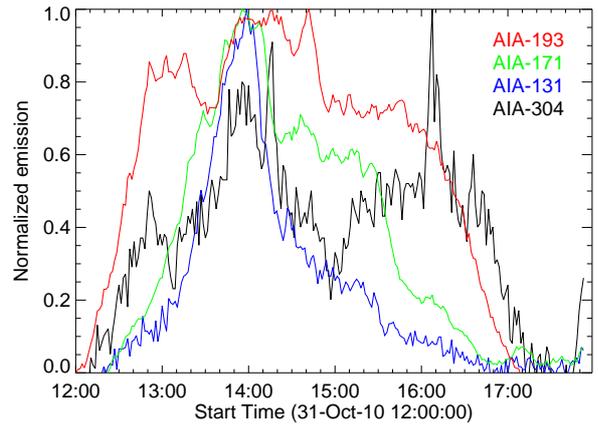}
 \caption{Normalized lightcurves of four AIA channels at
{\it A} in Fig. \ref{fig_limb}.}
 \label{fig_lc2}
\end{figure}

Apparent downflows are frequently observed in a time sequence
at temperatures lower than 1~MK.
The two righthand panels in Fig. \ref{fig_limb} show
the temporal variations of AIA-131\AA\, and 171\AA\,
between 13:15 UTC and 13:25 UTC,
at the beginning of a falling event.
A brightening in AIA-131\AA\,
and a darkening in the AIA-193\AA\,
are simultaneously detected at the cross signs {\it A} and {\it B}.

Figure \ref{fig_slice} presents a time distance plot,
where the two upper panels of Fig. \ref{fig_slice}
display original radiance in AIA-304\AA\, and 171\AA,
while the bottom panel shows a temporal variation of AIA-193\AA\,
to enhance small variations.
The brightening in AIA-171\AA\, began at 13:40 UTC
around a height of 80~Mm.
Small blobs in AIA-304\AA\, are seen to fall at a speed of 51~km~s$^{-1}$
starting 14:15 UTC.
A bright streak in AIA-171\AA\, appears at around 13:40 UTC and moves
down slowly until 14:40 UTC when it starts falling at a speed of
65~km~s$^{-1}$, following the cooler falling material seen in AIA-304\AA.
Just as the cool material reaches the surface, a noticeable upward motion
at a speed of 82~km~s$^{-1}$ in AIA-193\AA\, is initiated.
Similarly, in Fig. \ref{fig_slice_m},
an outward motion in AIA-193\AA\, is induced immediately following
the inward motion in AIA-171\AA.
The results indicate that the hot upflowing pattern is enhanced
after cool downflowing events.

Figure \ref{fig_lc2} displays normalized lightcurves
integrated over the 5\arcsec$\times$5\arcsec\, box
surrounding {\it A} in Fig \ref{fig_limb}.
The radiance at 17:30 UTC is subtracted from the measured radiance
to compensate for the background.
The assumption of a constant background is reasonable, since
the pre-event radiance at 12:00 UTC and
the postevent radiance at 17:30 UTC were quite similar.
AIA-193\AA\, shows a broad peak with a small dip at 13:30 UTC, when
AIA-131\AA\, and 171\AA\, rapidly increased.
This emission variation is interpreted as the cooling of plasma.
However, AIA-193\AA\, attained its maximum after
AIA-131\AA\, and 171\AA\, reached their peak emissions.
Then the AIA-131\AA\, and 171\AA\, decrease as the plasma
cooled and moved out of the box.
A small peak was noticed in AIA-304\AA\, at 14:15 UTC,
when falling blobs in AIA-304\AA\, started falling.
The AIA-335\AA\, only indicated a gradual change over several hours,
so it is not related to the falling event
(not shown in Fig.\ref{fig_lc2}).
The later AIA-304\AA\, peak at 16:08 UTC is due to another falling
event originating in the higher corona so is not directly
related to this cooling event.
In a six-hour movie in AIA-131\AA ,
eight sporadic downflowing events are identified in this
active region.
They seem to be seen in different loops rather than
repeatedly in the same loop.

\subsection{AIA response to cooling structures} \label{sect.model}

\begin{figure}
\centerline{\includegraphics[width=0.9\columnwidth]{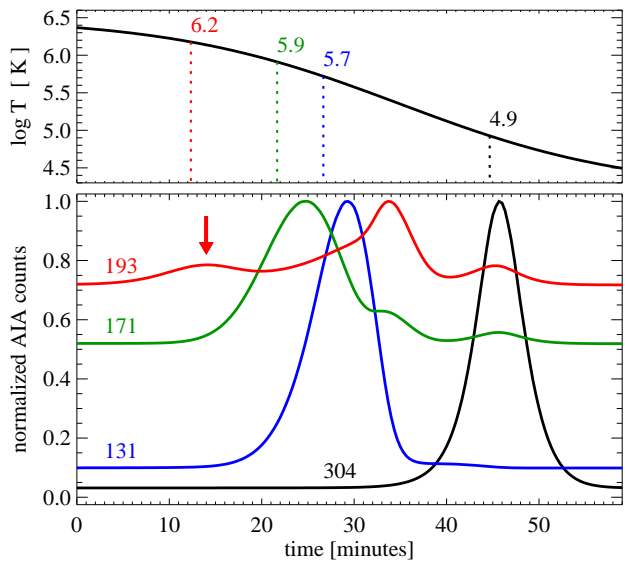}}
\caption{Emission for a model condensation at constant pressure ($n_{\rm e} = 3\times10^8 {\rm{cm}}^{-3}$ at $10^6$\,K).
The top panel shows the assumed temperature drop as a function of time.
The bottom panel displays the resulting emission as would be observed in four of the AIA bandpasses assuming a contribution of 1\,Mm along the line of sight for the condensation. A background corresponding to an active region emission measure distribution was added.
The vertical dotted lines indicate the times when the temperature of maximum response is reached for the four channels.
(See Sect.\,\ref{sect.model}).
\label{model}}
\end{figure}

When observing a cooling loop with AIA one would expect a successive brightening in the bandpasses according to the maximum of the temperature response functions, i.e. in \,the sequence AIA-193\AA, 171\AA, 131\AA, and 304\AA.
However, close inspection of the AIA response shows that the hot channels 211\AA\, and 193\AA\, show non-negligible contributions from cooler lines, e.g. there are numerous \ion{O}{v} lines in the 193\AA\, channel. This leads to a secondary maximum of the AIA-193\AA\, response at about ${\log}T{\approx}5.5$, which is about a factor of 10 weaker than the main peak at ${\log}T{\approx}6.2$ due to \ion{Fe}{xii}.

A loop that is cooling after the heating stopped will show a simultaneous decrease in temperature and pressure, and consequently the the AIA channels will indeed brighten according to their maximum response temperatures.
However, the situation is different for a loop that is undergoing catastrophic cooling \citep{schrijver2001}. In this case there is a loss of equilibrium between heating, radiative cooling, and heat conduction
\citep{mueller2003,mueller2004}, and condensation forms,
typically in the upper part of the loop. Through inflows from the lower parts of the loop, the pressure across this condensation remains roughly constant.
Thus, as the temperature drops, the density rises in the condensation. As the emission is proportional to the density squared, the contribution at low temperatures can even dominate the hot channels while the condensation is forming.

This is shown through a simple example in Fig.\,\ref{model}.
In a volume at constant pressure where the temperature drops gradually, multiple peaks can be found in some of the AIA bandpasses. In particular, the main peak of the hot 193\AA\, channel occurs \emph{after} the peaks of the cooler
171\AA\, and 131\AA\, channels. A smaller peak for the 193\AA\, channel can also be seen earlier (an arrow in Fig. \ref{model}), when the structure reaches the temperature of maximum response in the 193\AA\, channel. As mentioned above, that the 193\AA\, channel peaks after the cooler channels is due to the AIA temperature response functions, namely the contribution of cool lines to hot channels.

This gives a simple qualitative explanation of the peculiarities
of the AIA lightcurves shown in Fig. \ref{fig_lc2} and discussed
in Sect. 3.2.
However, the model does not perfectly agree with observations
because decay times of AIA-131\AA, 171\AA, and 193\AA\, lightcurves
are longer than the predictions.
This model is too simple to attempt a more quantitative comparison.

\section{Discussion}

AIA data in active regions reveal
brightness variations propagating both upward and downward.
In AIA-193\AA, the variations are always
present and propagate upward with the speeds of
57~km~s$^{-1}$ to 88~km~s$^{-1}$.
Our study from different viewing angles indicates that
these fluctuations originate at the footpoints of the loop system
and propagate along magnetic field lines diverging from the footpoint.
The blueshift observed at loop footpoints supports
an interpretation in terms of a constant hot upflow,
although the magnitude of the Doppler shift (up to $-20$~km~s$^{-1}$)
is smaller than the propagation speed along the investigated
cut in AIA-193\AA.
The low velocity could result from a superposition of different
velocity components under the spatial resolution of EIS.
It has been reported in hot emission lines (2~MK)
that the blue wing of the line profile is enhanced
in active regions, which is
interpreted as a signature of subsonic upflows
\citep{hara2008a,depontieu2009,peter2010}.
A detailed study of spectral profiles at the coronal loop
footpoints will be presented in a future paper.
We cannot rule out that the quasi-periodic intensity
fluctuation is caused by a propagating magneto-acoustic wave
(Fig.~\ref{fig_slice_m});
however, \citet{depontieu2010}
demonstrated that lightcurves from randomly driven upflows
and propagating waves are indistinguishable.

Sporadic downflows towards the loop footpoint are detected
in the cooler AIA-304\AA, 131\AA, and 171\AA\, channels,
which we interpret
as a cool plasma flowing down along the magnetic field lines.
The downflow of cool emission is reported
in previous studies \citep{schrijver2001, ugarte-urra2009}.
Our study revealed that the downflowing plasma is
initiated by a cooling in the corona.
The increase in AIA-171\AA\, and AIA-131\AA\, lightcurves coincide
with a small decrease in the AIA-193\AA\, lightcurve
(Fig. \ref{fig_lc2}).
The results support the idea of coronal condensation occurring
in the coronal loop \citep{mueller2003,mueller2004,mueller2005}.
However, AIA-193\AA\, increased again and remained high,
while AIA-131\AA\, decreased.
Although the observed sequence of peak emissions in multiple
AIA channels does not agree with the order of
maximum temperature response of the AIA channels,
it can be qualitatively explained by a cooling plasma
at a constant pressure (Fig. \ref{model}).
The simulated lightcurve of AIA-193\AA\, shows that the contribution of cool
emission lines within the broad passbands of the AIA channel is not
negligible in the event of coronal condensation.
Care should be taken when interpreting the lightcurves of AIA.
There is also a need for more realistic modeling of
cooling loops, which can be used to carry out a more quantitative
comparison with our observations.

The descent of cool plasma is seen in AIA-304\AA\, first,
and is followed by AIA-171\AA\, (Fig.\ref{fig_slice}).
It seems a cool dense core in AIA-304\AA\, is surrounded
by diffuse hotter plasma in AIA-171\AA.
This might be a reason for long decay times of lightcurves
from AIA-131\AA, 171\AA, and 193\AA.
It is interesting to note that the propagation pattern in
AIA-193\AA\, becomes more pronounced after a falling event in cool emission
(Figs. \ref{fig_slice_m} and \ref{fig_slice}).
A possible explanation is that the coronal loop is emptied
after the descent of cool plasma, so there is
plenty of room for hot upflows to propagate into the coronal loop.
This work demonstrates that a dynamic modeling
is essential for understanding the nature of coronal loops.

\begin{acknowledgements}
Authors are grateful to the SDO team for providing excellent data.
The German Data Center (GDC) for SDO is extensively used
in this work.
{\it Hinode} is a Japanese mission developed and launched by ISAS/JAXA,
with NAOJ as domestic partner and NASA and STFC (UK) as international
partners.
It is operated by these agencies in co-operation with ESA and NSC (Norway).
This work has been partly supported by WCU grant No. R31-10016
funded by the Korean Ministry of Education, Science, and Technology.
\end{acknowledgements}

\bibliographystyle{aa}
\bibliography{reference.bib}

\begin{thebibliography}{25}
\expandafter\ifx\csname natexlab\endcsname\relax\def\natexlab#1{#1}\fi

\bibitem[{{Antolin} {et~al.}(2010){Antolin}, {Shibata}, \&
  {Vissers}}]{antolin2010}
{Antolin}, P., {Shibata}, K., \& {Vissers}, G. 2010, \apj, 716, 154

\bibitem[{{Aschwanden} {et~al.}(2008){Aschwanden}, {Nitta}, {Wuelser}, \&
  {Lemen}}]{aschwanden2008c}
{Aschwanden}, M.~J., {Nitta}, N.~V., {Wuelser}, J., \& {Lemen}, J.~R. 2008,
  \apj, 680, 1477

\bibitem[{{Aschwanden} {et~al.}(2009){Aschwanden}, {Wuelser}, {Nitta}, {Lemen},
  \& {Sandman}}]{aschwanden2009}
{Aschwanden}, M.~J., {Wuelser}, J., {Nitta}, N.~V., {Lemen}, J.~R., \&
  {Sandman}, A. 2009, \apj, 695, 12

\bibitem[{Boerner {et~al.}(2011)Boerner, Edwards, Lemen, Rausch, Schrijver,
  Shine, Shing, Stern, Tarbell, Title, \& Wolfson}]{boerner2011}
Boerner, P., Edwards, C., Lemen, J., {et~al.} 2011, \solphys, submitted

\bibitem[{{Culhane} {et~al.}(2007){Culhane}, {Harra}, {James}, {Al-Janabi},
  {Bradley}, {Chaudry}, {Rees}, {Tandy}, {Thomas}, {Whillock}, {Winter},
  {Doschek}, {Korendyke}, {Brown}, {Myers}, {Mariska}, {Seely}, {Lang}, {Kent},
  {Shaughnessy}, {Young}, {Simnett}, {Castelli}, {Mahmoud}, {Mapson-Menard},
  {Probyn}, {Thomas}, {Davila}, {Dere}, {Windt}, {Shea}, {Hagood}, {Moye},
  {Hara}, {Watanabe}, {Matsuzaki}, {Kosugi}, {Hansteen}, \&
  {Wikstol}}]{culhane2007}
{Culhane}, J.~L., {Harra}, L.~K., {James}, A.~M., {et~al.} 2007, \solphys, 243,
  19

\bibitem[{{De Pontieu} \& {McIntosh}(2010)}]{depontieu2010}
{De Pontieu}, B. \& {McIntosh}, S.~W. 2010, \apj, 722, 1013

\bibitem[{{De Pontieu} {et~al.}(2009){De Pontieu}, {McIntosh}, {Hansteen}, \&
  {Schrijver}}]{depontieu2009}
{De Pontieu}, B., {McIntosh}, S.~W., {Hansteen}, V.~H., \& {Schrijver}, C.~J.
  2009, \apjl, 701, L1

\bibitem[{{Freeland} \& {Handy}(1998)}]{freeland1998}
{Freeland}, S.~L. \& {Handy}, B.~N. 1998, \solphys, 182, 497

\bibitem[{{Hara} {et~al.}(2008){Hara}, {Watanabe}, {Harra}, {Culhane}, {Young},
  {Mariska}, \& {Doschek}}]{hara2008a}
{Hara}, H., {Watanabe}, T., {Harra}, L.~K., {et~al.} 2008, \apjl, 678, L67

\bibitem[{{Kamio} {et~al.}(2010){Kamio}, {Hara}, {Watanabe}, {Fredvik}, \&
  {Hansteen}}]{kamio2010b}
{Kamio}, S., {Hara}, H., {Watanabe}, T., {Fredvik}, T., \& {Hansteen}, V.~H.
  2010, \solphys, 266, 209

\bibitem[{{Kjeldseth-Moe} \& {Brekke}(1998)}]{kjeldsethmoe1998}
{Kjeldseth-Moe}, O. \& {Brekke}, P. 1998, \solphys, 182, 73

\bibitem[{{Kosugi} {et~al.}(2007){Kosugi}, {Matsuzaki}, {Sakao}, {Shimizu},
  {Sone}, {Tachikawa}, {Hashimoto}, {Minesugi}, {Ohnishi}, {Yamada}, {Tsuneta},
  {Hara}, {Ichimoto}, {Suematsu}, {Shimojo}, {Watanabe}, {Shimada}, {Davis},
  {Hill}, {Owens}, {Title}, {Culhane}, {Harra}, {Doschek}, \&
  {Golub}}]{kosugi2007}
{Kosugi}, T., {Matsuzaki}, K., {Sakao}, T., {et~al.} 2007, \solphys, 243, 3

\bibitem[{Lemen {et~al.}(2011)Lemen, Title, Akin, Boerner, Drake, Duncan,
  Edwards, Friedlaender, Heyman, \& Hurlburt}]{lemen2011}
Lemen, J.~R., Title, A.~M., Akin, D.~J., {et~al.} 2011, \solphys, submitted

\bibitem[{{M{\"u}ller} {et~al.}(2005){M{\"u}ller}, {De Groof}, {Hansteen}, \&
  {Peter}}]{mueller2005}
{M{\"u}ller}, D.~A.~N., {De Groof}, A., {Hansteen}, V.~H., \& {Peter}, H. 2005,
  \aap, 436, 1067

\bibitem[{{M{\"u}ller} {et~al.}(2003){M{\"u}ller}, {Hansteen}, \&
  {Peter}}]{mueller2003}
{M{\"u}ller}, D.~A.~N., {Hansteen}, V.~H., \& {Peter}, H. 2003, \aap, 411, 605

\bibitem[{{M{\"u}ller} {et~al.}(2004){M{\"u}ller}, {Peter}, \&
  {Hansteen}}]{mueller2004}
{M{\"u}ller}, D.~A.~N., {Peter}, H., \& {Hansteen}, V.~H. 2004, \aap, 424, 289

\bibitem[{{O'Dwyer} {et~al.}(2011){O'Dwyer}, {Del Zanna}, {Mason}, {Sterling},
  {Tripathi}, \& {Young}}]{odwyer2011}
{O'Dwyer}, B., {Del Zanna}, G., {Mason}, H.~E., {et~al.} 2011, \aap, 525, A137

\bibitem[{{Peter}(2010)}]{peter2010}
{Peter}, H. 2010, \aap, 521, A51

\bibitem[{{Schrijver}(2001)}]{schrijver2001}
{Schrijver}, C.~J. 2001, \solphys, 198, 325

\bibitem[{{Tripathi} {et~al.}(2008){Tripathi}, {Mason}, {Young}, \& {Del
  Zanna}}]{tripathi2008}
{Tripathi}, D., {Mason}, H.~E., {Young}, P.~R., \& {Del Zanna}, G. 2008, \aap,
  481, L53

\bibitem[{{Ugarte-Urra} {et~al.}(2009){Ugarte-Urra}, {Warren}, \&
  {Brooks}}]{ugarte-urra2009}
{Ugarte-Urra}, I., {Warren}, H.~P., \& {Brooks}, D.~H. 2009, \apj, 695, 642

\bibitem[{{Warren} {et~al.}(2007){Warren}, {Ugarte-Urra}, {Brooks}, {Cirtain},
  {Williams}, \& {Hara}}]{warren2007}
{Warren}, H.~P., {Ugarte-Urra}, I., {Brooks}, D.~H., {et~al.} 2007, \pasj, 59,
  S675

\bibitem[{{Warren} \& {Winebarger}(2006)}]{warren2006}
{Warren}, H.~P. \& {Winebarger}, A.~R. 2006, \apj, 645, 711

\bibitem[{{Warren} {et~al.}(2003){Warren}, {Winebarger}, \&
  {Mariska}}]{warren2003}
{Warren}, H.~P., {Winebarger}, A.~R., \& {Mariska}, J.~T. 2003, \apj, 593, 1174

\bibitem[{{Winebarger} {et~al.}(2002){Winebarger}, {Warren}, {van
  Ballegooijen}, {DeLuca}, \& {Golub}}]{winebarger2002}
{Winebarger}, A.~R., {Warren}, H., {van Ballegooijen}, A., {DeLuca}, E.~E., \&
  {Golub}, L. 2002, \apjl, 567, L89

\end{thebibliography}

\end{document}